\documentclass[preprint,prd,aps,showpacs,showkeys,nofootinbib]{revtex4}

\def\XXint#1#2#3{{\setbox0=\hbox{$#1{#2#3}{\int}$}
     \vcenter{\hbox{$#2#3$}}\kern-.5\wd0}}

\usepackage{graphicx}
\usepackage{bm}
\usepackage{amssymb}
\usepackage{amsmath}
\usepackage{euscript}
\usepackage{enumitem}

\begin{document}

\title{The charge-carrier trapping effect on $1/f$ noise in monolayer graphene}

\author{K.~A.~Kazakov}
\author{T.~M.~Valitov}

\affiliation{Department of Theoretical Physics,
Physics Faculty,\\
Moscow State University, $119991$, Moscow, Russian Federation}

\begin{abstract}
The frequency exponent of $1/f$ noise in graphene--boron nitride heterostructures is known to have multiple extrema in its dependence on the charge carrier concentration. This behavior is explained in the present paper as a result of the charge carrier trapping by impurities in the boron nitride. A kinetic equation for the charge carriers subject to trapping and interacting with acoustic phonons is derived. This equation is solved numerically, and the equilibrium solutions are used to evaluate the frequency exponent according to the quantum theory of $1/f$ noise. It is found that the frequency exponent does develop several minima and maxima, provided that the trapping probability is sufficiently wide and has a threshold with respect to the charge carrier energy. A detailed comparison with the experimental data is made, and the results are used to estimate the energy threshold and the trapping cross-section.
\end{abstract}

\maketitle

\section{Introduction}

As all conducting materials, graphene exhibits the so-called $1/f$ noise on passing a constant electric current: at sufficiently small frequencies $f,$ the power spectrum $S(f)$ of voltage fluctuations measured across a graphene sample scales with frequency as
$S(f) \sim 1/f^{\gamma}\,,$ where the frequency exponent $\gamma$ is near unity. This noise has been detected in mono- as well as bi-layer graphene, suspended or immersed in liquid, and in various graphene-based heterostructures \cite{lin2008,pal2009,liu2009,xu2010,cheng2010,pal2011,kumar2015,kumar2016,kayyalha2015,karnatak2016,kakkar2020,das2021}. It possesses all characteristic properties of $1/f$ noise found in other materials, most notably the absence of a low-frequency flattening of $S(f),$ and the increase of its magnitude with decreasing sample size. The former property is intriguing from the theoretical standpoint as the divergence of $\int{\rm d}f S(f)$ for $\gamma\geqslant 1$ is apparently in conflict with the observed finiteness of the voltage fluctuation, while the latter is practically challenging in view of the numerous potential applications of graphene in nanotechnology.

The $1/f$ noise is generally considered as a result of fluctuations in the sample conductivity which are due to fluctuations either in the number of charge carriers or in their mobility. A problem with this approach is that it is manifestly inconsistent with the observed absence of a low-frequency spectrum flattening. In fact, $S(f)$ in this case necessarily flattens at $f\ll f_0,$ where $f_0$ is the inverse characteristic time of whatever underlying physical process. For conventional processes such as temperature fluctuations, generation-recombination of carriers in semiconductors, impurity migration, {\it etc.} the corresponding $f_0$s are all well above $1\,$Hz, whereas the existence of $1/f$-noise has been experimentally confirmed down to frequencies as low as $f=10^{-6.3}\,$Hz \cite{rollin1953,caloyannides}. Regarding the graphene-based structures, the electrical fluctuations are commonly related to the charge exchange between graphene and its environment, or to the charge rearrangements within the environment itself. Yet, the noise measurements in graphene reported in the literature do not probe the extremely low frequencies, usually covering only two-three decades above $1\,$Hz. Such measurements do not quite exclude the possibility that the spectrum might flatten at lower frequencies, and therefore, do not rule out the above-mentioned processes as possible mechanisms of noise generation. Another problem with the traditional approach is that no consistent theory based on the charge exchange has been developed so far which would derive from the first principles both the frequency profile and the magnitude of the noise power spectrum. For instance, a variant of this approach \cite{pal2011} involves four freely adjustable parameters including the noise magnitude.

On the other hand, there is a quite different mechanism of voltage fluctuations in materials possessing free-like charge carriers which has its roots in the quantum nature of the charge carrier interaction with the electromagnetic field. Namely, quantum indeterminacy in the values of voltage measured at different times sets a lower bound, $S_F(f),$ on the voltage power spectrum. The low-frequency asymptotic of $S_F(f)$ possesses all characteristic properties of the observed $1/f$ noise, and therefore is called the fundamental $1/f$ noise \cite{kazakov2020}. In particular, this theory naturally explains the absence of a low-frequency cutoff of  $1/f$ spectrum and reconciles it with finiteness of the total noise power. $S_F(f)$ depends on the sample geometry and on a number of parameters characterizing the energy-momentum dispersion of charge carriers and their interactions. In application to graphene and graphene-based heterostructures, $S_F(f)$ exhibits a curious feature of having pronounced peaks in its magnitude and in the value of the frequency exponent considered as functions of the charge carrier concentration. These peaks are centered at the Dirac point and are related to the masslessness of the charge carriers in graphene. A comparison with the experimental data on $1/f$ noise in single-layer graphene flakes and graphene--boron nitride heterostructures shows that within the accuracy of calculations, the shape of the peaks as well as the noise magnitude are adequately described by the theory.

At the same time, extensive measurements \cite{das2021} of the frequency exponent in graphene--hexagonal boron nitride (hBN) heterostructures provide also detailed data on its behavior away from the peak. Namely, the data reveals a certain non-monotonicity in the dependence of $\gamma$ on the charge carrier concentration which varies with the sample temperature. It was suggested in Ref.~\cite{kazakov2024} that the observed behavior of $\gamma$ can be a result of the charge carrier trapping by impurities in the boron nitride, and a toy model of this process was proposed to assess its effect on $\gamma.$ The model assumes that the charge carrier trapping rate is sufficiently low and uses a kinetic equation in which the collision integral is replaced by the simplest relaxation-type term to describe the charge carrier interaction with phonons. Though crude as it is, this model demonstrates that the frequency exponent does become non-monotonic on account of the charge-carrier trapping, provided that the trapping probability has an energy threshold.

The purpose of this paper is to obtain a quantitatively reliable description of the trapping effect on $\gamma.$ First of all, the leading (one-phonon) contribution to the charge carrier collision integral is carefully evaluated, and a kinetic equation for the charge-carriers subject to trapping is written down in Sec.~\ref{kineticequation}. A numerical procedure for solving this equation is described in Sec.~\ref{numerics}. General properties of the frequency exponent evaluated on equilibrium solutions of the kinetic equation are identified in Sec.~\ref{gammaproperties}. In Sec.~\ref{comparison}, these properties are used to qualitatively explain the observed behavior of $\gamma,$ and then a detailed quantitative comparison with the data from Ref.~\cite{das2021} is made. Section~\ref{conclusions} discusses the results and draws conclusions.

\section{Power spectrum of fundamental $1/f$ noise in graphene}

Consider a graphene sheet deposited on a substrate or encapsulated in a heterostructure which is part of the field-effect transistor to control the charge carrier density in graphene. Let the electrical voltage be measured between two electrodes in contact with graphene under a constant electric current. The electrodes may or may not coincide with the current-supplying leads. Then the power spectrum of fundamental voltage noise in graphene has the form
\begin{eqnarray}\label{fnoise}
S_F(f) = \frac{A U^2_0}{|f|}\int{\rm d}^2 \bm{q}\, \frac{\nu(\bm{q})}{|\bm{q}|}\left|\frac{f_*}{f}\right|^{\delta(\bm{q})},
\end{eqnarray}
\noindent where $U_0$ is the mean voltage between the electrodes,
$A$ is a combination of fundamental constants that depends also on the sample dimensions, $f_*$ is a frequency parameter dependent on the microscopic properties of the sample material, $\nu(\bm{q})\in [0,1]$ is the mean occupation number of the charge-carrier state with momentum $\bm{q}$ (and definite spin/pseudospin); $\delta(\bm{q})$ is a momentum-dependent contribution to the frequency exponent which in the case of piezoelectric interaction of charge carriers with the acoustic modes reads
\begin{eqnarray}\label{delta}
\delta(\bm{q}) = (2\nu(\bm{q})-1)\delta_m\,,
\end{eqnarray} where $\delta_m>0$ is a dimensionless parameter determined by the charge-carrier--phonon interaction. Explicit expressions for $A,\delta_m$ in terms of parameters characterizing the particle energy-momentum dispersion laws and their interactions are given in Ref.~\cite{kazakov2024}. They are not needed in the present consideration which is focused on how the frequency profile of $S_F$ is affected by the state filling; it will suffice to mention that $\delta_m$ is essentially the coupling constant of the charge-carrier--phonon interaction, and is typically in the range $0.2$--$0.4.$

It was found in Ref.~\cite{kazakov2024} that despite the momentum dependence of $\delta,$ $S_F(f)$ given by Eq.~(\ref{fnoise}) is perfectly well approximated by a power-law in the whole range of frequencies relevant in the $1/f$-noise studies:
\begin{eqnarray}\label{effectivegamma}
S_F(f) \sim 1/f^{\bar{\gamma}}\,, 
\end{eqnarray}
\noindent where $\bar{\gamma}$ is a number called effective frequency exponent. As defined, $\bar{\gamma}$ is a rather complicated functional of $\nu(\bm{q}),$ and therefore, it is affected by the processes changing the state filling, such as scattering and trapping of charge carriers. The aim of subsequent consideration is to quantitatively describe these effects on $\bar{\gamma}$ at varying charge carrier concentration.

\section{Kinetic equation for charge carriers subject to trapping}\label{kineticequation}

Evolution of the function $\nu(\bm{q})$ in the presence of traps is described by a kinetic equation of the form
\begin{eqnarray}\label{kineticeq}
\frac{{\rm d}\nu}{{\rm d}t} = {\rm St}[\nu,n] + {\rm Ts}[\nu],
\end{eqnarray}
\noindent where ${\rm St}$ is the charge-carrier collision integral, and ${\rm Ts}$ is the trap contribution to the rate of change of $\nu.$ The processes of trapping and detrapping of charge carriers drive them away from equilibrium, whereas particle collisions (this term is understood to include particle emission and absorption) act in the opposite direction. As usual, collisions are considered as instantaneous events between freely moving particles, according to which ${\rm St}[\nu,n]$ must vanish on the ideal equilibrium functions -- Fermi distribution for $\nu(\bm{q})$ and Planck distribution for the phonon mean number $n(\bm{k})$ [$\bm{k}$ is the phonon momentum]. On the contrary, charge carriers spend finite times in trap states, though the processes of trapping and detrapping themselves can be considered instantaneous. As a result, the right hand side of Eq.~(\ref{kineticeq}) no longer vanishes in the ideal equilibrium.

In the applications to be considered, the charge carrier concentration in graphene is comparatively small, so that the effect of collisions among charge carriers is negligible compared to that of the charge carrier interactions with phonons. Of major importance are the lowest-order processes, that is, processes involving one phonon. Their contribution to the collision integral in Eq.~(\ref{kineticeq}) will be found explicitly in Sec.~\ref{cintegral}. After that, the role of the multi-phonon precesses will be discussed and their contribution taken into account in a simplified manner. As to the kinetic equation for phonons, it is not needed in the considered approximation: a non-equilibrium correction to $n(\bm{k})$ in ${\rm St}[\nu,n]$ is a second-order effect with respect to the charge-carrier--phonon interaction. Less formally, because of the smallness of the charge carrier concentration, the charge-carrier--phonon interaction produces only a small deviation of $n(\bm{k})$ from the Planck distribution which is of minor importance compared to the effect of traps. Thus, the phonon distribution is taken henceforth to be
$$n(\bm{k}) = \left(\exp{\frac{u|\bm{k}|}{T}} - 1 \right)^{-1}\,,$$ where $T$ is the sample temperature and $u$ the acoustic velocity, assuming for simplicity the phonon energy-momentum dispersion isotropic.

The trapping term in Eq.~(\ref{kineticeq}) can be readily written down once the probability rate, $w(\bm{q}),$ of trapping the charge-carrier with momentum $\bm{q}$ is known. Assuming for simplicity the traps identical and denoting their mean occupation number by $\eta\in[0,1],$ the gain rate of charge carriers with momentum $\bm{q}$ is $\eta w(\bm{q})[1-\nu(\bm{q})],$ while the loss rate thereof is $(1 - \eta) w(\bm{q})\nu(\bm{q}).$ Thus,
\begin{eqnarray}\label{trapcontribution}
{\rm Ts}[\nu(\bm{q})] = \eta w(\bm{q})[1 - \nu(\bm{q})] - (1 - \eta) w(\bm{q})\nu(\bm{q}) = w(\bm{q})[\eta - \nu(\bm{q})].
\end{eqnarray}
\noindent It follows from the charge conservation that the rate of change of the mean trap occupation number satisfies \begin{eqnarray}\label{traprate}
N\frac{{\rm d}\eta}{{\rm d t}} = \int \frac{{\rm d}^2\bm{q}}{(2\pi\hbar)^2}\,w(\bm{q})[\nu(\bm{q}) - \eta],
\end{eqnarray}
\noindent where $N$ is the number of traps per unit graphene area. In the simplest case of traps acting independently, the trapping probability rate can be written as $w(\bm{q})=\sigma(\bm{q})v_F N,$ where $\sigma(\bm{q})$ is the 1D cross-section of trapping a charge carrier moving in the graphene sheet by a single trap in its surroundings, and $v_F$ is the Fermi velocity of charge carriers in graphene. In reality, however, dependence of $w(\bm{q})$ on the trap density can be very different from the simple proportionality because of the superposition of the trap fields, the possibility of charge-carrier hopping between traps, {\it etc.} No attempt will be made below to derive this function from the first principles. Instead, a simple model will be adopted in order to identify those characteristics of $w(\bm{q})$ that affect $\gamma$ most.

\subsection{The charge-carrier--phonon collision integral}\label{cintegral}

The one-phonon processes include emission and absorption of a phonon by a free-like charge carrier in states with definite (quasi)momentum along the graphene sheet. The probability rate of emission of a phonon with momentum $\bm{k}$ in the range ${\rm d}^2\bm{k}$ by a charge carrier with momentum $\bm{q}$ generally has the form
\begin{eqnarray}\label{rate}
\frac{b(\bm{k})}{|\bm{k}|}\delta(\bm{q}'+\bm{k}-\bm{q})
\delta(v_F |\bm{q}'| + u |\bm{k}| - v_F |\bm{q}|){\rm d}^2\bm{k}\,{\rm d}^2\bm{q}',
\end{eqnarray}
\noindent where momentum $\bm{q}'$ of the outgoing charge carrier belongs to the range ${\rm d}^2\bm{q}',$ and Dirac's delta-functions account for the total energy-momentum conservation. This expression is justified as follows. For sufficiently small charge carrier concentrations, $n \lesssim 10^{13}\,$cm$^{-2},$ Fermi momentum is small compared to the maximal momentum in the first Brillouin zone. Furthermore, in regard of the graphene-based heterostructures, of major importance are small-$\bm{k}$ acoustic phonons as they describe the long-range interactions of charge carriers propagating in the graphene layer with its surroundings. All the quasi-particles involved thus carry low momenta. Therefore, the conical energy-momentum dispersion holds for charge carriers as well as for phonons, and also the {\it Umklapp}-processes can be neglected. This yields the energy-momentum conservation as written above. Next, specific form of the coefficient function $b(\bm{k})$ is determined by the charge carrier interaction with the lattice vibrations. The factor $1/|\bm{k}|$ in Eq.~(\ref{rate}) is inserted because $b(\bm{k})=O(|\bm{k}|^0)$ when the charge-carrier--phonon interaction is generated by the sample piezoelectricity, that is, $b(\bm{k})$ depends only on the phonon momentum direction \cite{mahan1972}; it is customarily replaced by its angular average, $b_0,$ which will be assumed in what follows. The deformation potential adds to $b$ terms $O(\bm{k}^2),$ so that in general, $b(\bm{k}) = b_0 + b_2\bm{k}^2.$ But for the charge carrier concentrations of interest in graphene studies, contribution of the deformation potential can be neglected (provided that $b_0 \ne 0$). In fact, on dimensional grounds, $b_2/b_0$ is on the order of $(d/\hbar)^2,$ where $d$ is the lattice constant; therefore, $b(\bm{k})$ appreciably differs from $b_0$ only for $|\bm{q}|$s on the order of the maximal charge carrier momentum which are irrelevant at small charge carrier concentrations.

The rate of change in the number of charge carriers with momentum $\bm{q}$ due to the phonon emission is
\begin{eqnarray}
\int {\rm d}^2\bm{k}[n(\bm{k}) + 1]\frac{b_0}{|\bm{k}|}\Big\{&&\hspace{-0,3cm}[1 - \nu(\bm{q})]\nu(\bm{q}+\bm{k})\delta(v_F |\bm{q}| + u |\bm{k}| - v_F |\bm{q}+\bm{k}|) \nonumber\\&& - [1 - \nu(\bm{q}-\bm{k})]\nu(\bm{q})\delta(v_F |\bm{q}-\bm{k}| + u |\bm{k}| - v_F |\bm{q}|)
\Big\},\nonumber
\end{eqnarray}
\noindent where integration is over all $\bm{k}$ in the first Brillouin zone. Similarly, the rate of change in the number of charge carriers with momentum $\bm{q}$ due to phonon absorption is, by virtue of the reversibility principle,
\begin{eqnarray}
\int {\rm d}^2\bm{k}\,n(\bm{k})\frac{b_0}{|\bm{k}|}
\Big\{&&\hspace{-0,3cm}[1 - \nu(\bm{q})]\nu(\bm{q}-\bm{k})\delta(v_F |\bm{q}| - u |\bm{k}| - v_F |\bm{q}-\bm{k}|) \nonumber\\&& - [1 - \nu(\bm{q}+\bm{k})]\nu(\bm{q})\delta(v_F |\bm{q}+\bm{k}| - u |\bm{k}| - v_F |\bm{q}|)
\Big\}.\nonumber
\end{eqnarray}
\noindent
Adding these rates and making the change of integration variable $\bm{k}\to -\bm{k},$ the collision integral thus takes the form
\begin{eqnarray}\label{collisionintegral}
&&{\rm St}[\nu(\bm{q}),n] = \int \frac{{\rm d}^2\bm{k}\, b_0}{|\bm{k}|}\nonumber\\&&
\times\Big\{\big[\nu(\bm{q}+\bm{k})(1 - \nu (\bm{q}) + n(\bm{k})) - \nu(\bm{q})n(\bm{k})\big]\delta(v_F |\bm{q}+\bm{k}| - v_F |\bm{q}| - u |\bm{k}|) \nonumber\\&& + \big[\nu(\bm{q}+\bm{k})n(\bm{k}) - \nu(\bm{q})(1 - \nu(\bm{q}+\bm{k}) + n(\bm{k}))\big]\delta(v_F |\bm{q}+\bm{k}| - v_F |\bm{q}| + u |\bm{k}|)
\Big\}.
\end{eqnarray}
\noindent For the present purposes it will suffice to determine  $\nu(\bm{q})$ neglecting the external electric field; dependence of $S_F(f)$ on this field is basically described by the factor $U^2_0$ in Eq.~(\ref{fnoise}), whereas the indirect effect via $\nu(\bm{q})$ becomes significant only at extremely large field strengths comparable to the interatomic. By this reason, ${\rm d}\nu/{\rm d t}$ in Eq.~(\ref{kineticeq}) can be replaced by $\partial \nu/\partial t.$
Since all the coefficient functions in the kinetic equation are now direction-independent, its solution $\nu(\bm{q})$ is also a function of $|\bm{q}|\equiv q$ only. 

Using the smallness of the acoustic velocity in comparison with Fermi velocity, $u/v_F \approx 10^{-2},$ the collision integral can be simplified as follows. With the understanding that all the quantities involved are functions of the momenta moduli, we change their notation to $\nu(q),n(k),$ {\it etc.,} and use the energy conservation to write
$$\nu(|\bm{q}+\bm{k}|) = \nu\left(q \pm \frac{u}{v_F}k\right) = \nu(q) \pm \nu^{\prime}(q)\frac{u}{v_F}k + \frac{1}{2}\nu''(q)\left(\frac{u}{v_F}k\right)^2 + O(k^3),$$ where $\nu^{\prime}(q)\equiv \partial \nu(q)/\partial q,$ and the plus (minus) sign refers to the first (second) term in braces in Eq.~(\ref{collisionintegral}). Upon integration over directions of $\bm{k},$ the collision integral becomes, in the leading order with respect to $u/v_F$ (it is to be noted that $n(k)= O\left(1/u\right)$ for $k\ll T/u$),
\begin{eqnarray}
{\rm St}[\nu(q),n] =&& \int\limits_{0}^{2q} {\rm d}k\,\frac{b_0u}{qv^2_F}\left[1 - \left(\frac{k}{2q}\right)^2\right]^{-1/2}\nonumber\\&& \times
\left\{q\nu^{\prime}(q)[1 - 2\nu(q)] + q\nu''(q)\frac{u}{v_F}k n(k) + 2\nu^{\prime}(q)\frac{u}{v_F}kn(k) + 2\nu(q)[1 - \nu(q)]\right\}.\nonumber
\end{eqnarray}
\noindent The charge carrier concentrations in graphene studies are typically $n\lesssim 10^{13}\,$cm$^{-2},$ which corresponds to Fermi energy $\varepsilon_F\lesssim 0.3\,$eV. Therefore, at not too low temperatures, $T\gg 40\,$K, one has $uk\lesssim
(u/v_F)\varepsilon_F\ll T,$ so that the phonon distribution is well approximated by $n(k)=T/(uk),$ and the $k$-integration is then easily done:
\begin{eqnarray}\label{collisionintegralexp}
{\rm St}[\nu(q),n] =&& \frac{\pi u b_0}{v^2_F}\left\{q\nu^{\prime}(q)[1 - 2\nu(q)] + q\nu''(q)\frac{T}{v_F} + 2\nu^{\prime}(q)\frac{T}{v_F} + 2\nu(q)[1 - \nu(q)]\right\} \nonumber\\&& \equiv \frac{1}{t_0 q}\frac{\partial}{\partial q}q^2\left\{\frac{\partial\nu(q)}{\partial q}\frac{T}{v_F} + \nu(q)[1 - \nu(q)]\right\}, \quad t_0 \equiv \frac{v^2_F}{\pi u b_0}\,.
\end{eqnarray}
\noindent It is straightforward to check that this collision integral is nullified by Fermi distribution $$\nu_0(q) = \left(\exp{\frac{v_F q - \mu}{T}} + 1 \right)^{-1}\,,$$ $\mu$ denoting the charge-carrier chemical potential. There is also a continuum of parasitic equilibria satisfying $$\nu(1 - \nu) + \frac{\partial\nu}{\partial q}\frac{T}{v_F} = \frac{c}{q^2}$$ with arbitrary constant $c\ne 0.$ These must be ruled out by imposing some condition on solutions of the kinetic equation. We will use the requirement that $\nu(q)$ be exponentially small at large $q.$ In other words, the large-momentum tail of $\nu(q)$ is required to coincide with that of $\nu_0(q).$ This condition means physically that the statistics of charge carriers with sufficiently high energy is not affected by the traps.

As was evident in advance from the smallness of energy exchange in the charge-carrier--phonon interactions, Eq.~(\ref{collisionintegralexp}) describes diffusion of the charge carriers with respect to energy, though unlike the classical effect, this diffusion is nonlinear. In addition to this leading term, the charge carrier collision integral also receives contributions from other processes, such as the charge carrier scattering on impurities, their direct Coulomb scattering, {\it etc.} Despite the relative smallness of these contributions, they must be explicitly included to provide a large-time saturation of the diffusion process, which otherwise would continue indefinitely. As this is their sole purpose, specific form of these contributions is immaterial, and they will be collectively taken into account by adding to the right hand side of Eq.~(\ref{collisionintegralexp}) a relaxation-type term $(\nu_0 - \nu)/t_*,$ $t_*$ denoting a characteristic time of the mentioned processes. This brings the kinetic equation to the form
\begin{eqnarray}\label{kineticeqfinal}
\frac{\partial\nu(q)}{\partial t} = \frac{1}{t_0 q}\frac{\partial}{\partial q}q^2\left\{\frac{\partial\nu(q)}{\partial q}\frac{T}{v_F} + \nu(q)[1 - \nu(q)]\right\} + w(q)[\eta - \nu(q)] + \frac{\nu_0 - \nu}{t_*}\,.
\end{eqnarray}
\noindent

\section{Evaluation of the frequency exponent}

\subsection{Transition to natural units}

In order to solve Eq.~(\ref{kineticeqfinal}) numerically it is convenient to introduce dimensionless momentum $\varkappa$ and time $\tau$ according to
\begin{eqnarray}\label{dimensionless}
q = \varkappa \frac{T}{v_F}\,, \quad t = \tau t_0\,.
\end{eqnarray}
\noindent Equation (\ref{kineticeqfinal}) then becomes
\begin{eqnarray}\label{kineticeqfinaldless}
\frac{\partial\nu}{\partial \tau} = \frac{1}{\varkappa}\frac{\partial}{\partial \varkappa}\varkappa^2\left\{\frac{\partial\nu}{\partial \varkappa} + \nu(1 - \nu)\right\} + W(\varkappa)(\eta - \nu) + \frac{\nu_0 - \nu}{\tau_*},
\end{eqnarray}
\noindent where
\begin{eqnarray}\label{coeffuncdless}
W(\varkappa) \equiv t_0 w\left(\varkappa \frac{T}{v_F}\right)\,, \quad \tau_* \equiv \frac{t_*}{t_0}\,.
\end{eqnarray}
\noindent Together with the expression for the rate of change of $\eta$ following from Eq.~(\ref{traprate}),
\begin{eqnarray}\label{trapratedimless}
\frac{{\rm d}\eta}{{\rm d \tau}} = \frac{T^2}{2\pi N\hbar^2v^2_F}\int {\rm d}\varkappa\,\varkappa\,W(\varkappa)(\nu - \eta),
\end{eqnarray}
\noindent Eq.~(\ref{kineticeqfinaldless}) determines the charge carrier distribution over momenta. It is to be noted that while ${\rm d}\eta/{\rm d\tau}$ explicitly depends on $N,$ steady solutions do not, the dependence of $\nu,\eta$ on $N$ being only implicit via that of $W(\varkappa).$ The total concentration of electric charges, $n,$ which is the sum of free charge carriers in graphene, $n_0,$ and the trapped ones, does depend on $N$ explicitly:
\begin{eqnarray}\label{totalconc}
n = n_0 + 4\eta N, \quad n_0 = 4\int \frac{{\rm d}^2\bm{q}}{(2\pi\hbar)^2}\,\nu = \frac{2T^2}{\pi \hbar^2 v^2_F}\int {\rm d}\varkappa\,\varkappa \nu,
\end{eqnarray}
\noindent where the factor of four accounts for the charge carrier spin and pseudospin.
As for the trapping probability distribution, we adopt the simplest rectangular distribution
\begin{eqnarray}\label{trapprobability}
W(\varkappa) = W_0\theta(\varkappa - \varkappa_1)\theta(\varkappa_2 - \varkappa),
\end{eqnarray}
\noindent where the amplitude $W_0$ and the edges $\varkappa_{1,2}$ are free parameters, and $\theta(x)$ is the Heaviside function. As the subsequent calculations show, this model not only qualitatively reproduces basic features of the observed frequency exponent, but also allows a sufficiently accurate quantitative approximation thereof.

\subsection{Numerical solutions of the kinetic equation}\label{numerics}

\begin{figure}[bp]
\includegraphics[width=7.5cm]{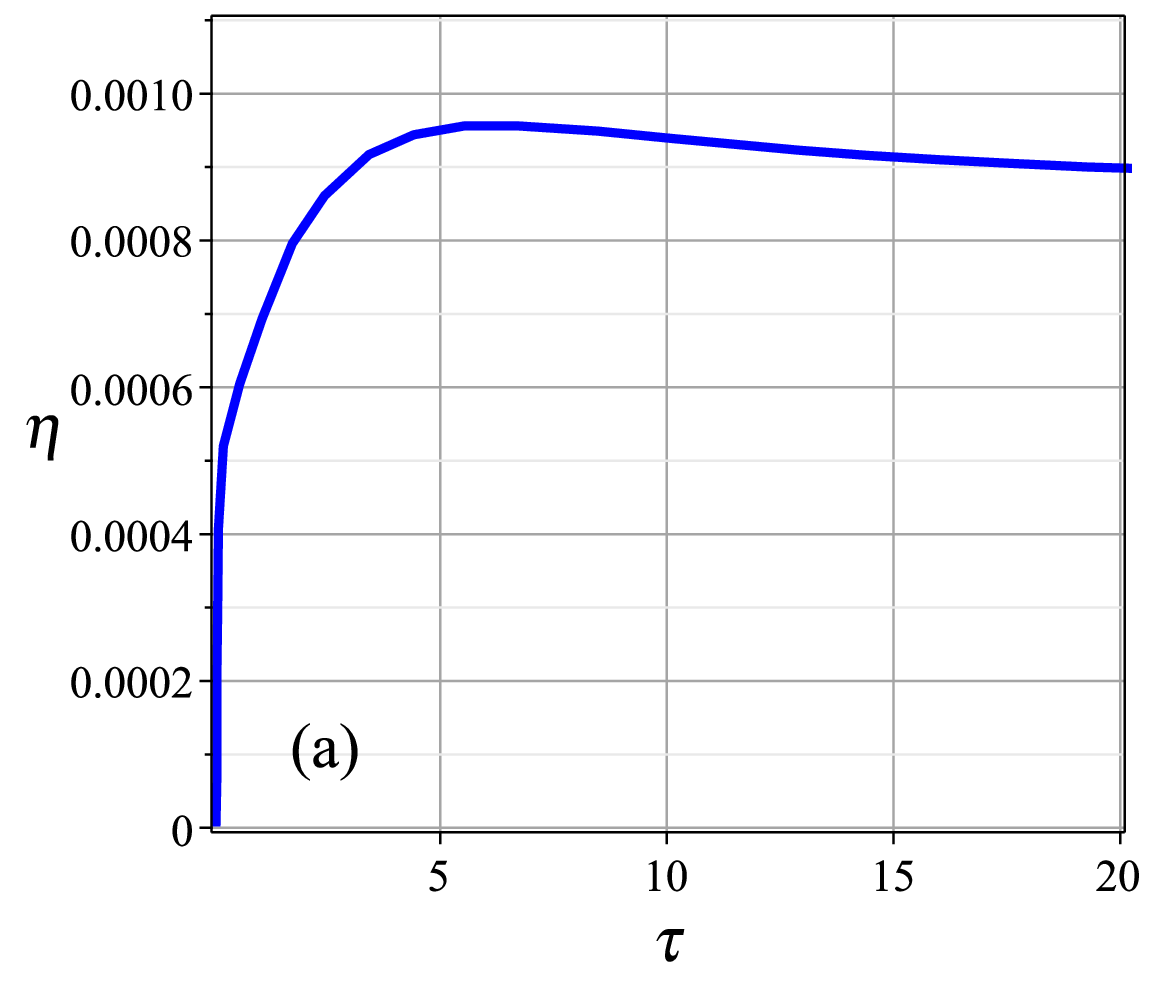}
\includegraphics[width=7.5cm]{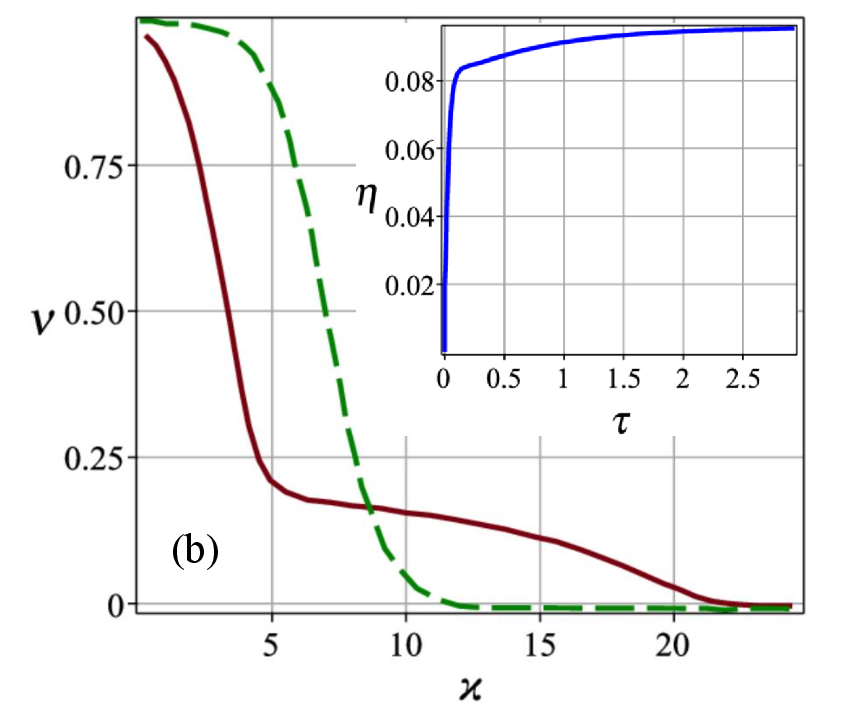}
\caption{(a) Evolution of the mean trap occupation number for $W_0=1,$ $\varkappa_1=4,$ $\varkappa_2 = 30,$ $m=1,$ $\tau = 10.$ (b) Initial (dash) and equilibrium (solid) charge carrier momentum distribution for $W_0 = 5,$ $\varkappa_1=4,$ $\varkappa_2 = 20,$ $m=7,$ $\tau = 10^3.$ The corresponding evolution of the trap occupation number is shown in the inset.}\label{fig1}
\end{figure}

Equations (\ref{kineticeqfinaldless}), (\ref{trapratedimless}) are numerically integrated with the following initial condition at $\tau=0$:
\begin{eqnarray}\label{initialcond}&&
\eta = 0, \quad \nu = \nu_0 = \left(\exp[\varkappa - m] + 1\right)^{-1}, \quad m \equiv \frac{\mu}{T}\,.
\end{eqnarray}
\noindent An explicit finite-difference scheme is used in which the $\varkappa$-derivatives of $\nu$ are approximated to the fourth order in $\Delta\varkappa,$ where the momentum grid spacing $\Delta\varkappa$ is typically $0.1.$ The $\tau$-derivatives are approximated by the forward differences to successively compute $\nu,\eta$ with a step $\Delta\tau$ sufficiently small to ensure the scheme stability ($\Delta\tau = 10^{-5} - 10^{-4}$). Regarding the requirement of exponential smallness of $\nu(\varkappa)$ at large $\varkappa,$ it is supposed to be effected by the relaxation term in the kinetic equation: since all other terms on the right of this equation vanish as $\nu$ becomes nearly constant at large $\varkappa>\varkappa_2,$ one has $\nu = \nu_0$ there. However, in view of the relative smallness of the factor $1/\tau_*,$ this mechanism may fail at large times, resulting in a loss of equilibrium. Therefore, to enforce the equality $\nu = \nu_0$ at large $\varkappa,$ we treat $\tau_*$ as $\varkappa$-dependent (as it actually is), being equal to the characteristic time of non-diffusive processes at $\varkappa<\tilde{\varkappa},$ but taking on a much smaller value otherwise. In subsequent computations, we set $\tilde{\varkappa} = 30$ and $\tau_*(\varkappa>\tilde{\varkappa}) = 10^4.$ The notation $\tau_*$ is kept for its value at $\varkappa<\tilde{\varkappa}.$ 

A typical evolution of the mean trap occupation number is shown in Fig.~\ref{fig1}(a). $\eta(\tau)$ is often non-monotonic, and after a steep, nearly vertical initial rise on the interval $0<\tau\lesssim 0.1,$ it evolves slowly and may have a local minimum, a maximum, or neither. During the rise, charge carriers partially fill in the initially empty traps, whereas the subsequent evolution on $0.1<\tau \ll \tau_*$ is driven by the nonlinear diffusion until an equilibrium is reached at times of a few of $\tau_*.$ Calculations show that for $\Delta\tau = 10^{-5},$ this takes $10^5-10^6$ time steps. An important general property of $\eta$ is that its equilibrium value rapidly decreases as $(\varkappa_2-\varkappa_1)$ grows. For $(\varkappa_2-\varkappa_1) = 1$ to $10$ and $m\lesssim 10,$ $\eta$ is comparable to unity, but for $(\varkappa_2-\varkappa_1)\approx 30,$ $\eta$ takes on values $10^{-5}-10^{-4}$ at $m\approx -2$ to $10^{-3}-10^{-2}$ at $m\approx 10.$

Regarding the influence of traps on the charge carrier momentum distribution, it is most important that the traps effectively change the relative amount of charge carriers occupying states with small $\nu,$ Fig.~\ref{fig1}(b). As is seen from Eqs.~(\ref{fnoise}), (\ref{delta}), the more such carriers in the sample, the lower the effective frequency exponent. To put it most plainly, whether $\bar{\gamma}$ is larger than unity depends on the relative amounts of the charge carriers occupying states with $\nu > 1/2$ and $\nu < 1/2.$ Of course, a precise value of $\bar{\gamma}$ can be obtained only by numerical evaluation of the integral in Eq.~(\ref{fnoise}) on a given solution of the kinetic equation. Technically, $\bar{\gamma}$ is found simply as $\log \left[S_F\left(1\,{\rm Hz}\right)/S_F\left(10^3\,{\rm Hz}\right)\right]/3.$

Numerical analysis reveals a remarkable fact that $\bar{\gamma}$ is in general a non-monotonic function of the charge carrier concentration. Naturally, this non-monotonicity occurs at sufficiently large values of the magnitude $W_0$ of the trapping probability and sufficiently wide range $[\varkappa_1,\varkappa_2]$ of the charge carrier momenta subject to trapping.

\subsection{General properties of the function $\gamma(n)$}\label{gammaproperties}

\begin{figure}[bp]
\includegraphics[width=7.5cm]{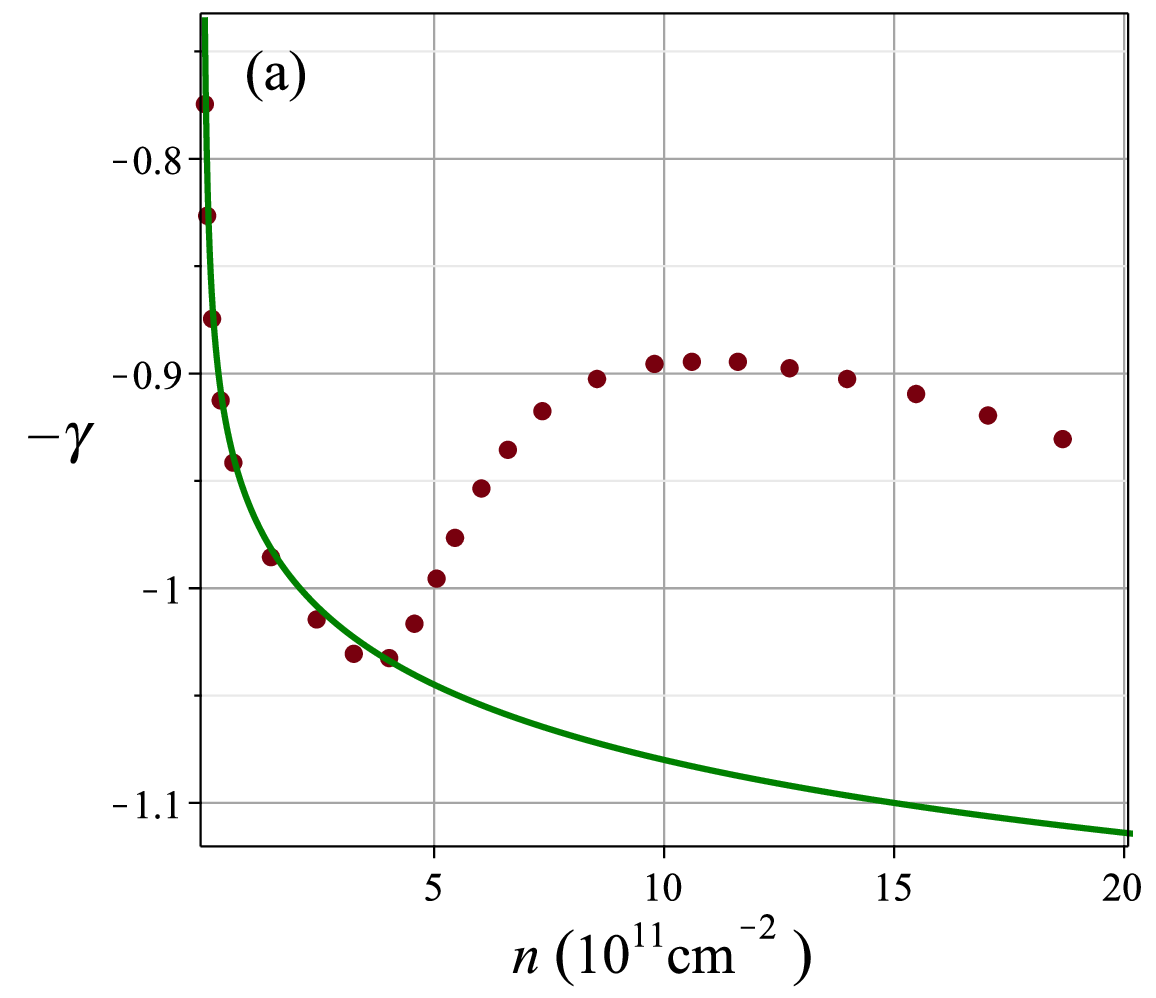}
\includegraphics[width=7.5cm]{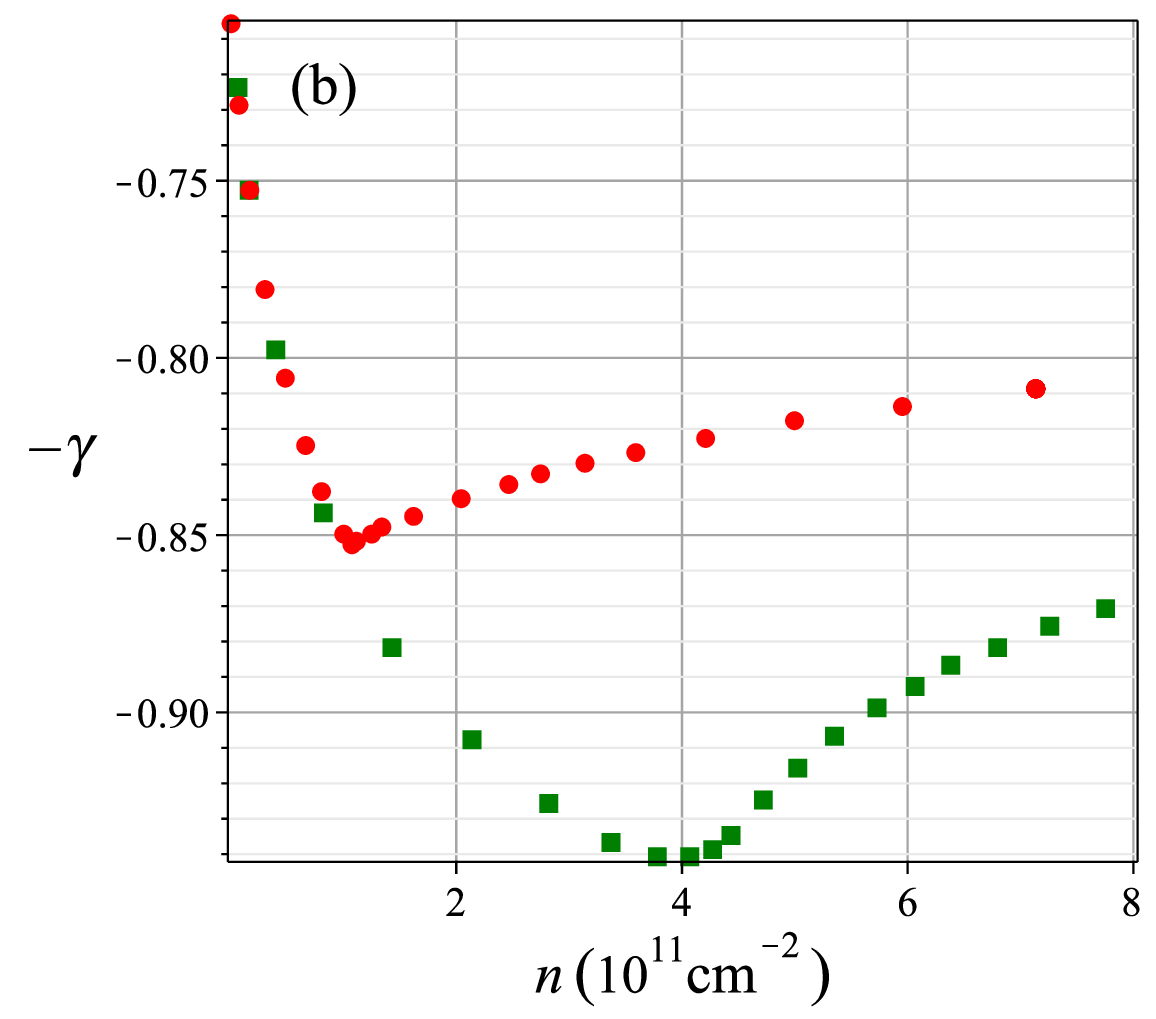}
\includegraphics[width=7.5cm]{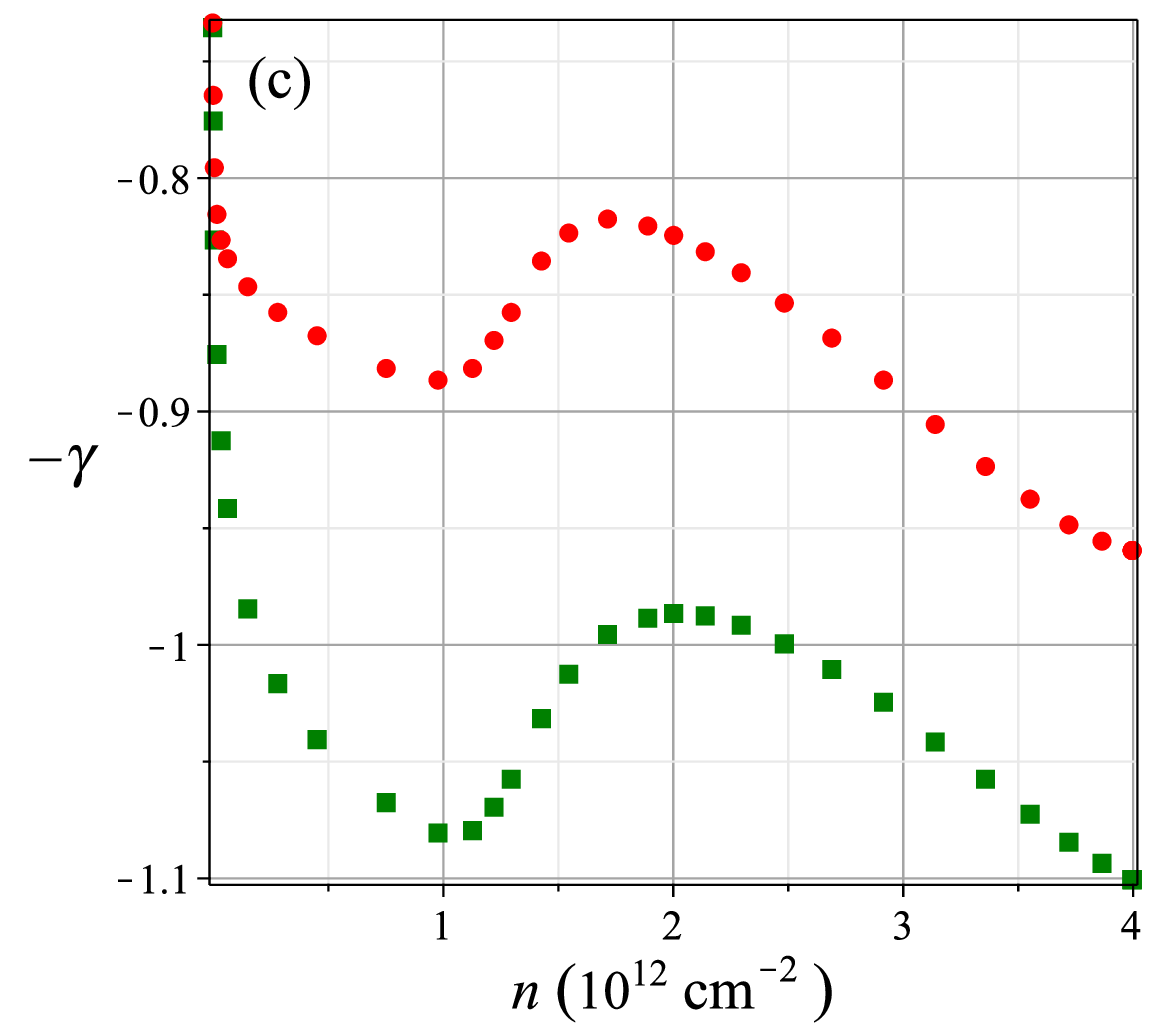}
\includegraphics[width=7.5cm]{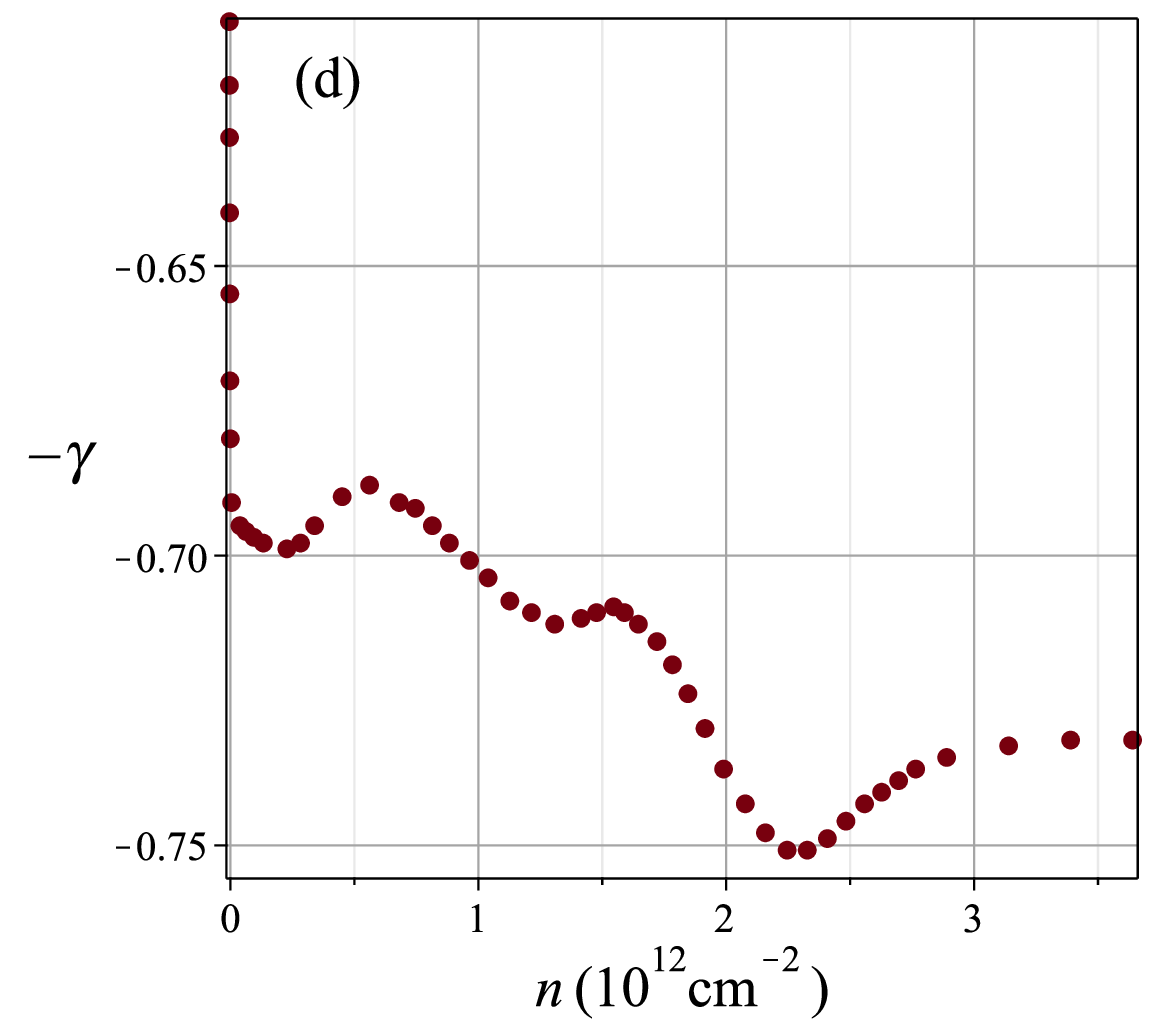}
\caption{Effective frequency exponent versus charge carrier concentration at statistical equilibrium. (a) Trap-free dependence at $T=100\,$K, $\delta_m=0.3,$ $f_* = 0.1\,$Hz (solid curve); effect of traps with $W_0 = 2,$ $\varkappa_1=10,$ $\varkappa_2=80,$ $\tau_* = 10$ (marks). (b) Effect of increasing $\tau_*$ at $T=300\,$K, $\delta_m=0.3,$ $W_0=10,$ $\varkappa_1=5,$ $\varkappa_2=80,$ $f_* = 0.1\,$Hz: $\tau_* = 10$ (boxes), $\tau_* = 100$ (circles). (c) Effect of decreasing $f_*$ at $T=100\,$K, $\delta_m=0.3,$ $W_0=1,$ $\varkappa_1=15,$ $\varkappa_2=80,$ $\tau_* = 10$: $f_* = 0.1\,$Hz (boxes), $f_* = 0.001\,$Hz (circles). (d) A multi-extrema pattern; $T=80\,$K, $W_0=1,$ $\varkappa_1=11,$ $\varkappa_2=80,$ $\delta=0.4,$ $f_*=10^{-4}\,$Hz, $\tau_*= 1.$}\label{fig2}
\end{figure}
\noindent

We now summarize qualitative properties of the frequency exponent dependence on the charge carrier concentration, $n,$ in the state of equilibrium. Having in mind comparison with the experiment \cite{das2021}, they are formulated in terms of $(-\bar{\gamma}).$
\begin{enumerate}[label={\bf T\arabic*}]
    \item{At small $n,$ such that $\nu(q) \ll 1$ for all $q,$ $\delta(\bm{q})\approx - \delta_m$ [Cf. Eq.~(\ref{delta})], therefore, $\bar{\gamma} = 1 - \delta_m.$ As $n$ is increased, the mean occupation numbers of states with small $q$s rapidly grow and reach unity. The growth is rapid because the density of charge carrier states vanishes at zero energy. If, as is usually done, the electron (hole) regime of conduction corresponds to positive (negative) values of $n,$ then the graph of $(-\bar{\gamma}(n))$ will have a pronounced peak at its centre called the charge neutrality point (CNP), Fig~\ref{fig2}(a).}
  \item{In the case of ideal charge carrier distribution (that is, in the absence of traps), $(-\bar{\gamma})$ continues to monotonically decrease as $n$ is further increased. This is because charge carriers tend to fill states with larger momenta only after all states with smaller momenta have been filled almost completely. The relative amount of charge carriers occupying the states with small $\nu$ thus gradually decreases, and so $\bar{\gamma}$ asymptotically tends to the value $(1 + \delta_m),$ Fig.~\ref{fig2}(a) (solid line).}
  \item{Switching on the traps results in a redistribution of charge carriers over momenta. The states that were almost empty in the absence of traps become partially filled, provided that the probability of trapping with the given momentum is finite. It readily follows from Eq.~(\ref{kineticeqfinaldless}) that for sufficiently high trapping rate, $\nu(q)$ will be nearly constant equal to $\eta$ over the range $[\varkappa_1, \varkappa_2]$ of momenta where the traps are active. This outflow of charge carriers from states with $\nu\approx 1$ to states with smaller $\nu$s leads to increase of $(-\bar{\gamma}).$ Thus, at sufficiently high trapping rates, the graph $(-\bar{\gamma}(n))$ develops a local minimum, Fig.~\ref{fig2}(a) (marks). Location of the minimum is strongly correlated with the value of the trapping probability threshold: its abscissa rapidly shifts to larger $|n|$ as $\varkappa_1$ increases.}
  \item{Location of the minimum is also affected by the value of $\tau_*,$ but the major effect of this parameter is on the shape of the minimum: as $\tau_* \gg 1$ is increased, the graph $(-\bar{\gamma}(n))$ becomes more narrow (``sharper'') in a vicinity of the minimum, Fig.~\ref{fig2}(b).}
  \item{The rise of $(-\bar{\gamma}(n))$ away from the local minimum stops at some point, because as $n$ grows, the contributions of states with $\nu(q)\approx 1$ eventually prevail. Thus, the local minimum is followed by a local maximum.}
  \item{Pass the local maximum, $\bar{\gamma}(n)$ usually monotonically approaches its asymptotic value $(1+\delta_m),$ though in general its structure is rather complicated, and before reaching the value $(1+\delta_m)$ it may have several maxima and minima, Fig.~\ref{fig2}(d). Extrema proliferate on increasing $\varkappa_1$ and decreasing $\tau_*.$ In view of {\bf T3}, this means that for a given charge carrier concentration, decreasing sample temperature adds new extrema.}
  \item{The global structure of $(-\bar{\gamma}(n))$ is noticeably affected by the frequency parameter $f_*$ appearing in Eq.~(\ref{fnoise}). This parameter is related to the charge-carrier self-energy, and its value essentially depends on the form of the charge-carrier and phonon energy-momentum dispersion laws in the entire Brillouin zone \cite{kazakov2024}. As such, it is treated as a free parameter within the present theory based on the conical approximations of these laws which are valid only sufficiently close to CNP. Numerical solutions show that as $f_*$ is decreased, the graph of $(-\bar{\gamma}(n))$ is retracted upwards, its central peak remaining fixed, the first local minimum retaining its horizontal position, and the first local maximum shifting to smaller $n$s, Fig.~\ref{fig2}(c).}
\end{enumerate}
\noindent

It is worth recalling that the described behavior of the frequency exponent is for the simplest trapping probability distribution. More general distributions will produce more complicated patterns.

\section{Comparison with the experiment}\label{comparison}

The frequency exponent $\gamma$ appearing in the empirical law $S(f)\sim 1/f^{\gamma}$ is clearly an important parameter whose measurement may help identify the noise origin in a given system. However, such measurements are often largely scattered, which is probably why the experimentalists commonly quote only typical values of the frequency exponent, without revealing trends in its behavior under varying physical conditions. Moreover, its deviations from unity are often discarded in order to approximate the power spectrum by the simplest formula $S(f)=\alpha_H U^2_0/N_0f,$ $N_0$ being the total number of charge carriers in a sample \cite{hooge1969}. This is despite the fact that doing so affects {\it dimensionality} of the Hooge's ``constant'' $\alpha_H,$ distorting thereby its physical meaning. The work \cite{das2021} is exceptional in this respect. It provides extensive data on the frequency exponent measured under varying gate voltage in field-effect transistors based on a graphene-hBN heterostructure. The channel in this case is a single-layer graphene encapsulated in hBN. This structure is known for its low noise levels compared to devices fabricated on silicon dioxide \cite{kayyalha2015,stolyarov2015}. The analysis carried out in Ref.~\cite{kazakov2024} has shown that the level of noise measured in Ref.~\cite{das2021} is near the quantum bound, Eq.~(\ref{fnoise}), and that the noise is to be attributed to  out-of-plane piezoelectric activity in the considered heterostructure. This activity is induced in graphene by carbon impurities and other defects in boron nitride. Substitution of the boron and nitrogen atoms by carbon creates trapped states for both electrons and holes. In these circumstances, given that the noise level is near the quantum bound, it is natural to ask whether it is possible to explain, on the basis of Eq.~(\ref{fnoise}), the intricate behavior of the frequency exponent observed in Ref.~\cite{das2021}. The subsequent consideration explores this possibility within the trapping model developed in Secs.~\ref{kineticequation}, \ref{numerics}.

\begin{figure}[bp]
\includegraphics[width=10cm]{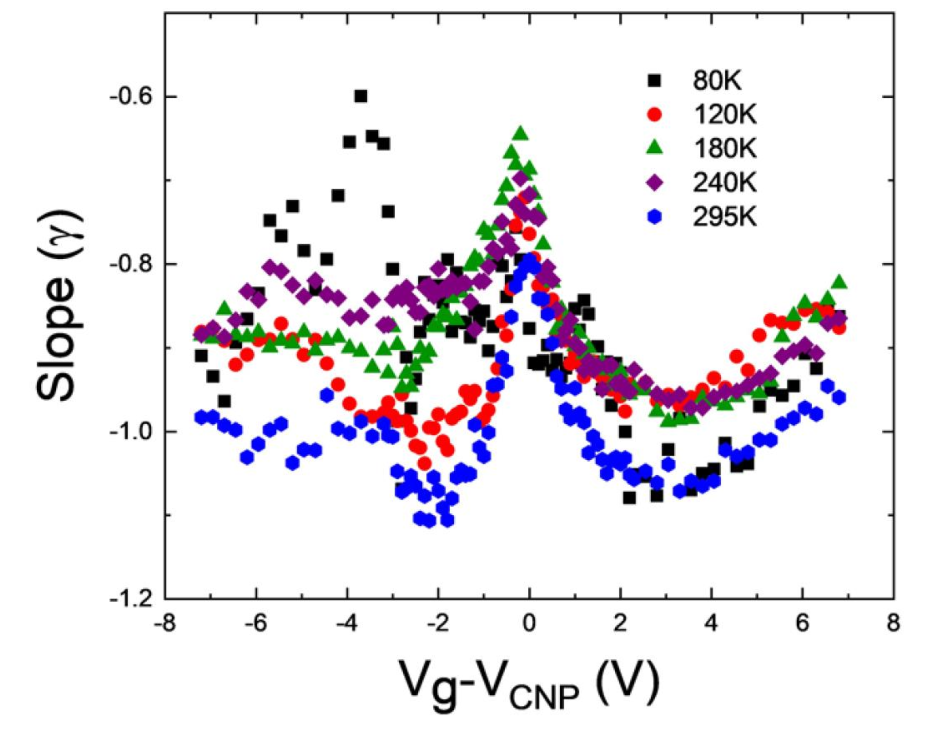}
\caption{Frequency exponent versus gate voltage as measured in Ref.~\cite{das2021} at various temperatures in a hBN-graphene-hBN heterostructure. The ordinate is {\it minus} the frequency exponent, $(-\gamma)$ (A.~K.~Behera, private communication). Reprinted with permission from A.~K.~Behera {\it et al.}, {\it ACS Applied Electronic Materials} {\bf 3}(9), 4126 (2021). Copyright \copyright  2021, American Chemical Society.}\label{das}
\end{figure}

Experimental data obtained in Ref.~\cite{das2021} for the frequency exponent is reproduced in Fig.~\ref{das}. It is seen that despite a considerable spread, the data reveals certain trends in the dependence of $(-\gamma)$ on the voltage $(V_g - V_{CNP}) \equiv V,$ where $V_g$ is the gate voltage and $V_{CNP}$ its value at CNP:

\begin{enumerate}[label={\bf E\arabic*}]
  \item{Each set of experimental marks representing $(-\gamma(V))$ at given temperature exhibits a sharp peak at CNP.}
  \item{Away from CNP, this peak goes over into a local minimum on both electronic and hole conduction regions ($e$- and $h$-regions, for brevity), except probably the $h$-region at $T=240\,$K.}
  \item{In some cases, notably at lower temperatures, the local minimum of $(-\gamma(V))$ is followed by a local maximum further away from CNP.}
    \item{The marks representing $(-\gamma(V))$ at $T=295\,$K are distinctly vertically offset relative to the rest of the data.}
\end{enumerate}

\begin{figure}[bp]
\includegraphics[width=7.5cm]{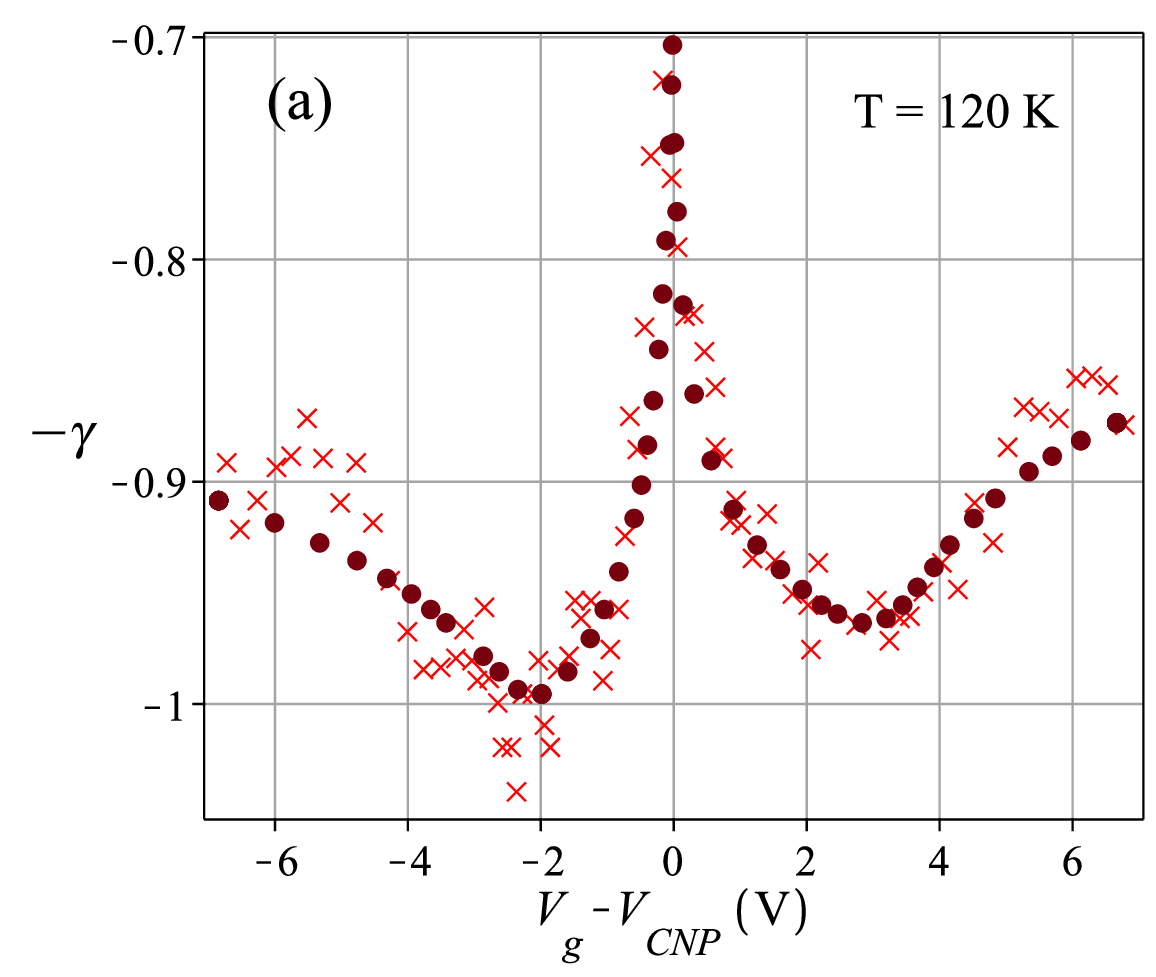}
\includegraphics[width=7.5cm]{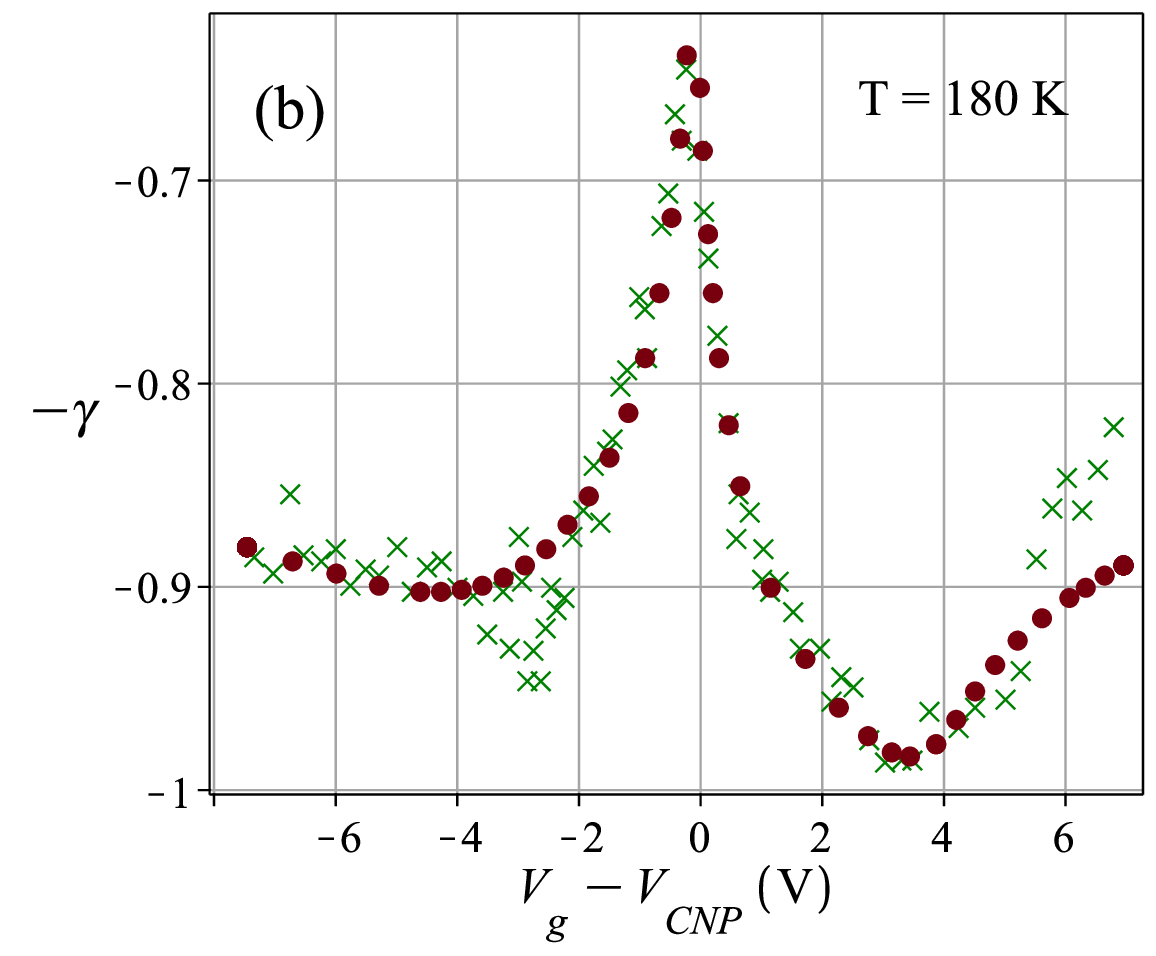}
\caption{The frequency exponent as measured in graphene/hBN heterostructure \cite{das2021} (crosses) and calculated using Eqs.~(\ref{fnoise}), (\ref{delta}) on solutions of the kinetic equation (\ref{kineticeqfinaldless}) (circles). (a) $T=120\,$K; $e$-region: $W_0=3,$ $\varkappa_1 = 7.7,$ $\delta_m = 0.28,$ $f_*=0.04\,$Hz, $\tau_* = 50$; $h$-region: $W_0=3,$ $\varkappa_1 = 7.2,$ $\delta_m = 0.30,$ $f_*=0.3\,$Hz, $\tau_* = 100.$ (b) $T=180\,$K; $e$-region: $W_0=6,$ $\varkappa_1 = 5.5,$ $\delta_m = 0.35,$ $f_*=0.5\,$Hz, $\tau_* = 10$; $h$-region: $W_0=1,$ $\varkappa_1 = 4.5,$ $\delta_m = 0.5,$ $f_*=0.8\,$Hz $\tau_* = 10.$}\label{t120}
\end{figure}

\begin{figure}[bp]
\includegraphics[width=7.5cm]{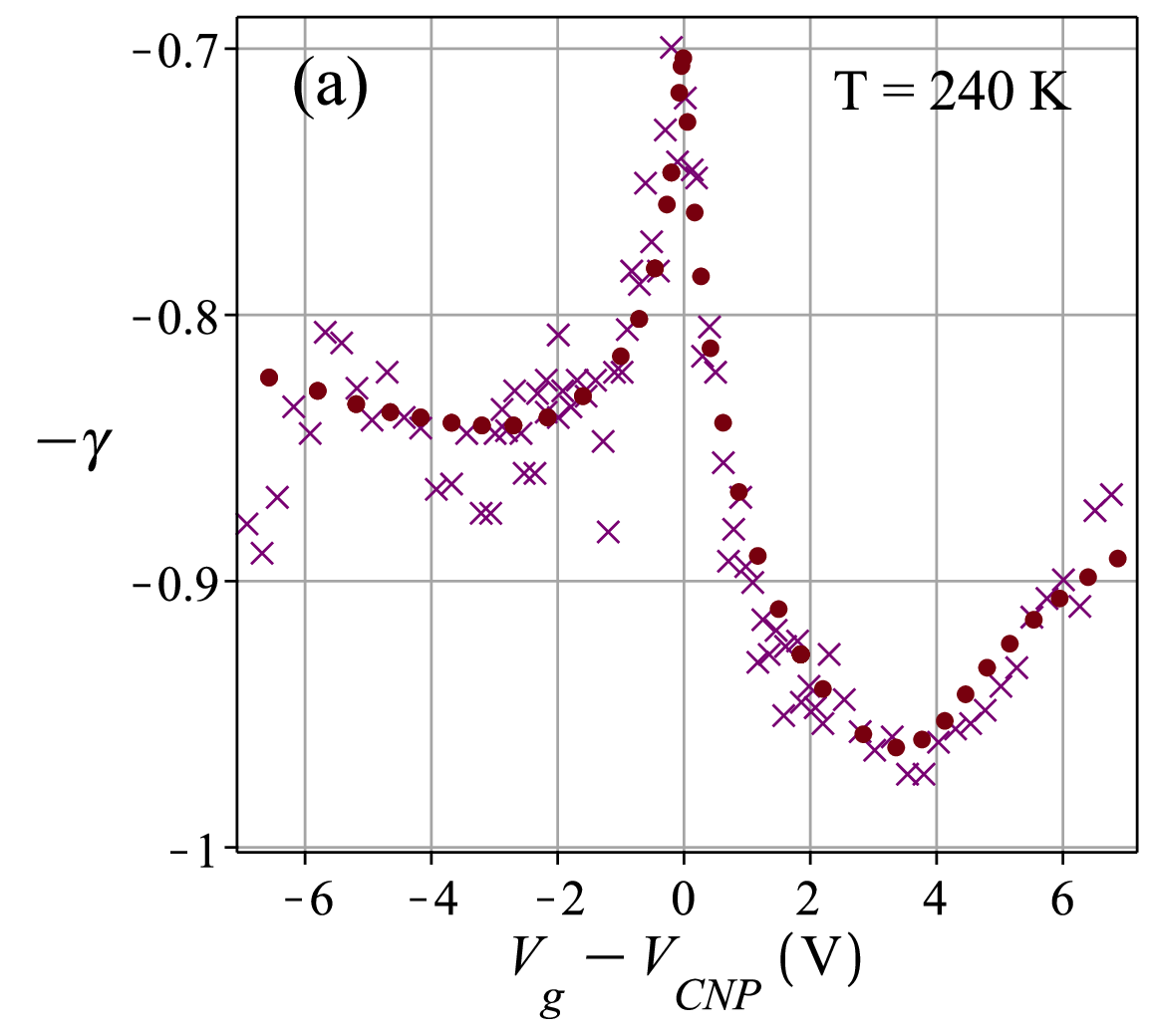}
\includegraphics[width=7.75cm]{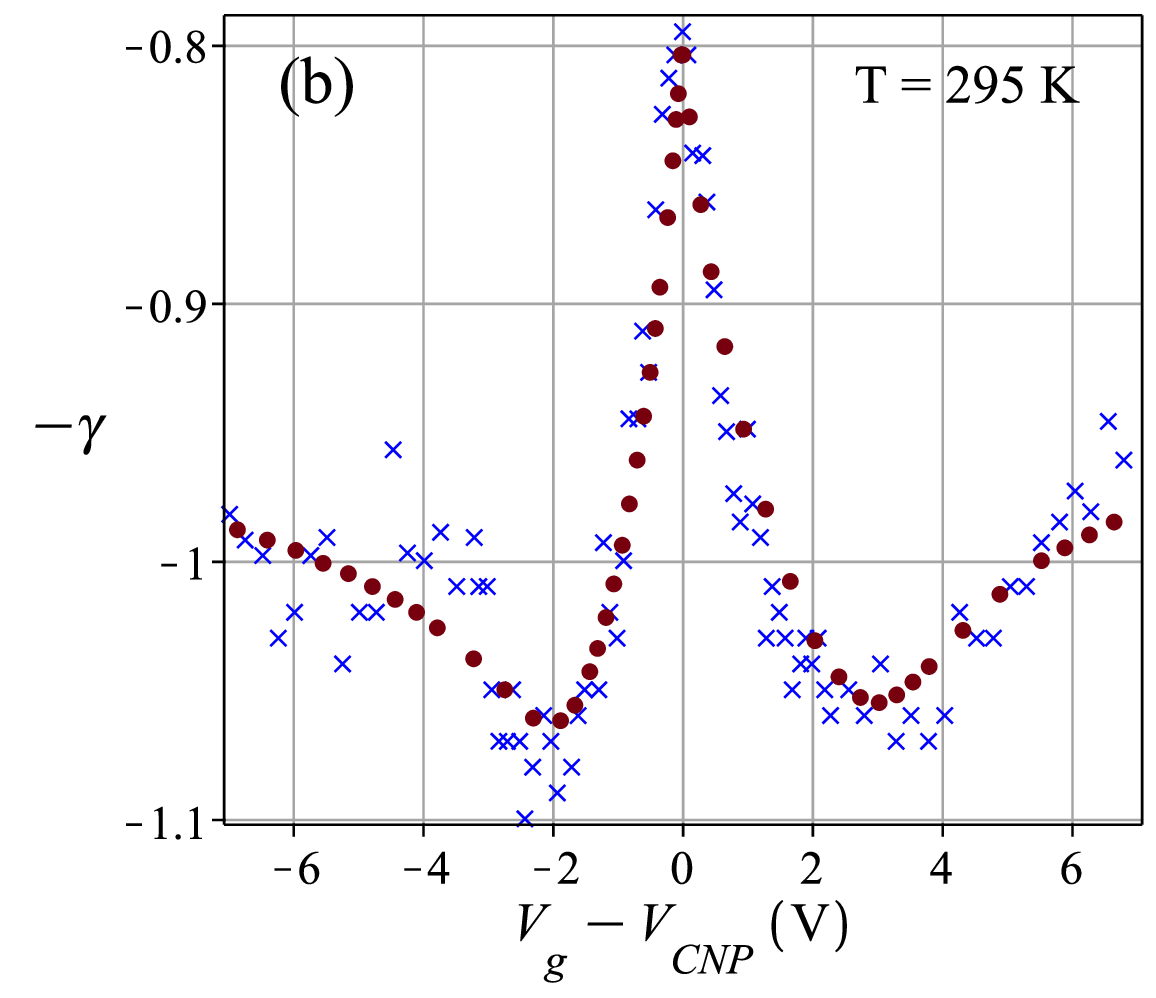}
\caption{Same as in Fig.~\ref{t120} for (a) $T=240\,$K; $e$-region: $W_0=8,$ $\varkappa_1 = 4,$ $\delta_m = 0.30,$  $f_*=0.25\,$Hz, $\tau_* = 4$; $h$-region: $W_0=1,$ $\varkappa_1 = 4.75,$ $\delta_m = 0.30,$ $f_*=0.005\,$Hz, $\tau_* = 200.$ (b) $T=295\,$K; $e$-region: $W_0=200,$ $\varkappa_1 = 2.25,$ $\delta_m = 0.2,$ $f_*=0.3\,$Hz, $\tau_* = 0.1$; $h$-region: $W_0=200,$ $\varkappa_1 = 1.8,$ $\delta_m = 0.2,$ $f_*=1.5\,$Hz, $\tau_* = 0.1.$}\label{t240}
\end{figure}

Taking into account that the charge carrier concentration is linearly proportional to $V,$ one readily sees that observations {\bf E1}, {\bf E2}, and {\bf E3} are qualitatively explained respectively by properties {\bf T1}, {\bf T3}, and {\bf T5} of the function $(-\bar{\gamma}(n))$ evaluated on solutions of the kinetic equation. We now proceed to a quantitative comparison of  theoretical and experimental graphs of the frequency exponent. To this end, we use an estimate $C \approx 10^{-4}\,$F/m given by the authors of Ref.~\cite{das2021} for the capacitance of the studied heterostructure. This relates the total charge concentration $n$ to $V,$ as $V = en/C \approx 1.6\times10^{-11}n,$ where $e$ is the elementary charge, and it is assumed that $V$ is measured in volts, and $n,$ in inverse centimeters squared. As to the trap concentration, we use throughout the value $N=10^{12}\,$cm$^{-2}$ which is a typical impurity concentration for substrated graphene. It is to be noted though that because of the smallness of equilibrium $\eta,$ a precise value of $N$ is actually immaterial for computing the total charge concentration as long as $N \lesssim 10^{13}\,$cm$^{-2}.$

Consider first the case of the sample temperature $T = 120\,$K. Figure \ref{t120}(a) shows an approximation of the experimental data (crosses) by the theory (circles). Since $\bar{\gamma}$ is a rather complicated functional of the trapping probability distribution, it is difficult to develop a regular minimization procedure for approximating the data. By this reason, all approximations presented below are obtained by trial and error until the fit satisfies the eye. We find, in particular, that in the $e$-region, $\varkappa_1 = 8,$ $f_*=0.03\,$Hz, whereas in the $h$-region,  $\varkappa_1 = 7,$ $f_*=0.2\,$Hz. The difference in $\varkappa_1$ accounts for the different abscissas of local minima according to {\bf T3}, and that in $f_*,$ for the vertical graph extension according to {\bf T7}. As to parameter $\varkappa_2,$ calculations show that it has nearly no effect on $\bar{\gamma}$ as long as it is sufficiently large. Therefore, we henceforth set $\varkappa_2 = 80$ in all cases. It is seen that $(-\gamma(V))$ is well approximated by the theory all the way from its central peak through the local minima, but the theoretical $(-\gamma(V))$ is somewhat less steep than the experimental near the local maxima which are apparent at $|V_g - V_{CNP}|\approx 6\,$V. Similar observations apply to the case of $T = 180\,$K, Fig.~\ref{t120}(b), except that the local maxima on the experimental $(-\gamma(V))$ are not seen. Also, the large experimental spread near $V_g - V_{CNP} \approx - 3\,$V does not allow one to clearly identify position of the local minimum. This spread persists in the $h$-region also at higher temperatures, Fig.~\ref{t240}. In the $e$-region, on the other hand, things are quite similar to those in Fig.~\ref{t120}. Thus, the four figures consistently demonstrate that near the local maxima, the theoretical graphs are less steep than the experimental, which  suggests that the theoretical maxima are located at larger $|V|$s, as is the case at $T=120\,$K. Numerical analysis shows that the ratio of the local maximum's abscissa to that of the local minimum decreases as $\varkappa_1$ is increased. Therefore, the found mismatch may potentially be caused by a uniform overestimation of the calculated $V$ or $n,$ that is, by possibly underestimated capacitance $C$ or overestimated factor $(T/\hbar v_F)^2$ in Eq.~(\ref{totalconc}). Specifically, the latter strongly depends on Fermi velocity for which we use an approximate value of $10^8\,$cm/s. Calculations show that already a moderate increase of $v_F$ noticeably changes the relative position of the extrema, making it more like that in Fig.~\ref{fig2}(c).

Next, comparison of the numerical values of the basic parameters given in Figs.~\ref{t120}, \ref{t240} reveals that the case $T=295\,$K is sharply distinct from the others by the values of $W_0$ and $\tau_*,$ both in $e$- and $h$-regions. At the same time, the variation of other parameters with temperature is more gradual and moderate. Therefore, observation {\bf E4} signifies a sharp enhancement of non-diffusive processes near the room temperature. In particular, it indicates occurrence of resonant contributions to the trapping cross-section, as discussed further below in Sec.~\ref{conclusions}.

\begin{figure}[bp]
\includegraphics[width=10cm]{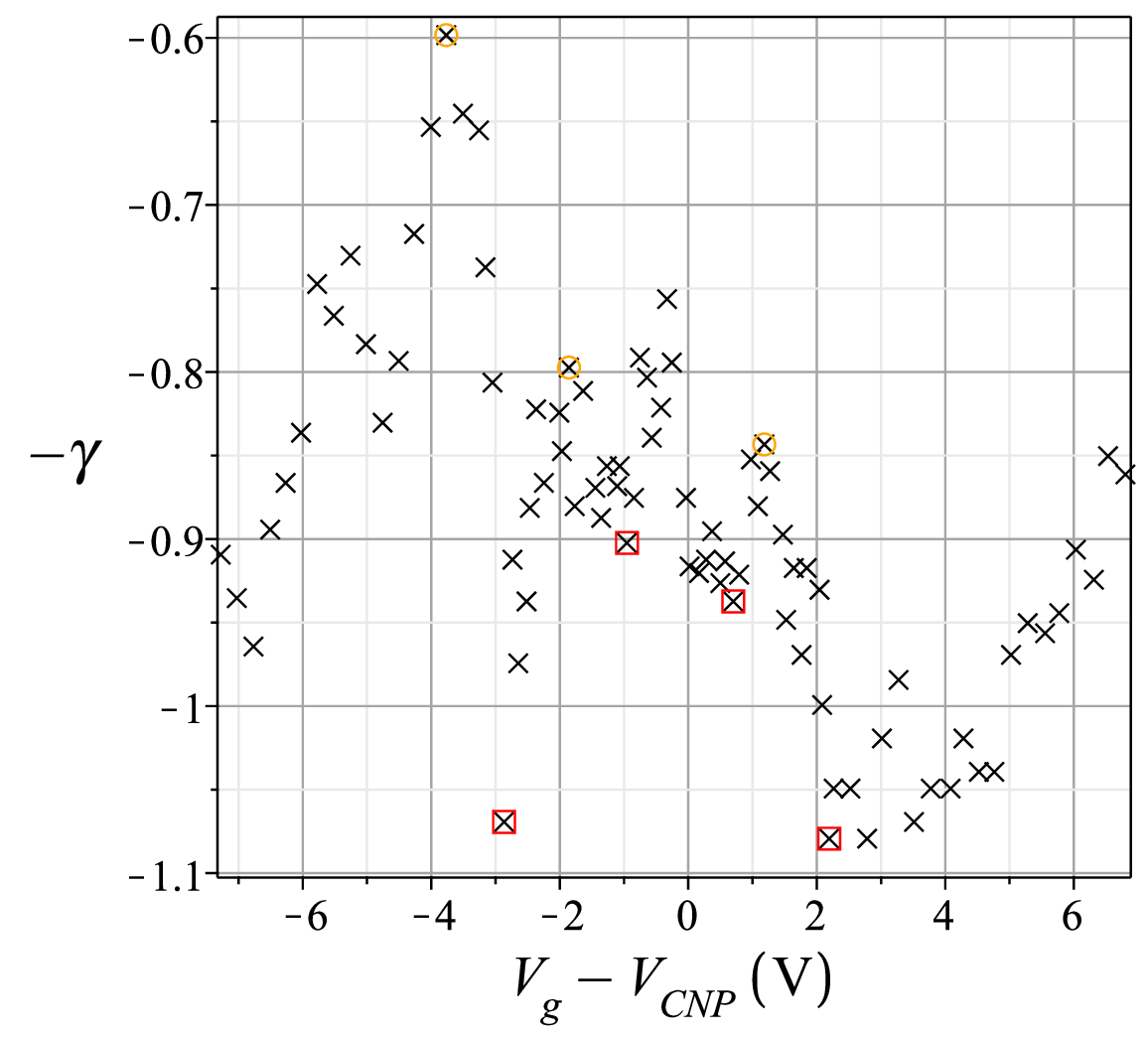}
\caption{Frequency exponent versus gate voltage at $T=80\,$K as read off from Fig.~\ref{das}. Suspected maxima (minima) are enclosed in circles (boxes).}\label{das80}
\end{figure}

At last, regarding the case $T = 80\,$K, we do not attempt to approximate the experimental data because of the too large scatter which makes it difficult to reliably identify local extrema, Fig~\ref{das80}. We only note that there seem to be present multiple minima and maxima in each of the two conduction regions, a possibility mentioned in {\bf T6}.

\section{Discussion and conclusions}\label{conclusions}

The frequency exponent characterising $1/f$ noise in graphene-hBN heterostructures was observed in Ref.~\cite{das2021} to exhibit rather intricate patterns in its dependence on the charge carrier concentration. In order to explain them, we derived a kinetic equation for the charge carriers subject to trapping by impurities in the heterostructure. In the model proposed, evolution of the charge carrier distribution is driven by their interaction with phonons and traps, all other processes being lumped together by a relaxation-type term in the kinetic equation. By solving this equation numerically and evaluating the frequency exponent on  equilibrium solutions according to Eq.~(\ref{fnoise}), we reproduced all of its observed properties, most notably, the occurrence of local extrema of $\gamma(V).$ In this computation, the simplest rectangular distribution of the trapping probability was assumed which not only allowed of a qualitative explanation, but also provided a satisfactory quantitative approximation of the experimental data. A general conclusion of our analysis is that the frequency exponent is significantly affected by the traps already when the mean trap occupation numbers are as small as $10^{-2}.$ Thus, the trapping-detrapping processes result in a considerable redistribution of charge carriers over momenta despite the fact that for typical impurity concentrations, trapped at any instant is only a small fraction of charge carriers.

The obtained results also allow one to draw some conclusions regarding the trapping mechanism. Namely, for the occurrence of side extrema of the function $\bar{\gamma}(n),$ it is essential that the distribution of trapping probability over charge carrier momentum (or, equivalently, its energy) be sufficiently wide and have a momentum threshold. The quantitative comparison of Sec.~\ref{comparison} shows that the dimensionless threshold drops from $\varkappa_1 \approx 7$ to $\varkappa_1 \approx 2$ as the sample temperature is increased from $120\,$K to $295\,$K. Since $\varkappa = \varepsilon/T,$ Cf. Eq.~(\ref{dimensionless}), it follows that the dimensionful threshold belongs to a significantly narrower interval approximately $0.05\,$eV to $0.08\,$eV. Next, to assess strength of the charge carrier interaction with impurities, one can estimate the trapping cross-section using the simplified picture of independent traps, wherein $w =\sigma v_F N.$ Combining this formula with Eq.~(\ref{coeffuncdless}) and the definition of $\delta_m$ from Ref.~\cite{kazakov2024} gives $$\sigma = W_0\delta_m \pi^2\frac{u}{v_F}\frac{1}{aN},$$ where $a = 70\,$nm is the heterostructure thickness. With $N = 10^{12}\,$cm$^{-2}$ for the impurity concentration, this gives $\sigma \approx W_0\delta_m 10^{-8}\,$cm, suggesting non-resonant reactions in the case of $W_0 \lesssim 10.$ Referring back to Figs.~\ref{t120}, \ref{t240}, this takes place at all temperatures except $295\,$K. On the other hand, a much larger value of $W_0$ at the room temperature indicates that the interaction of charge carriers with traps becomes resonant. These observations suggest the existence of a series of localized states of charge carriers with energies $\gtrsim 0.05\,$eV and significantly different lifetimes.

It should be mentioned that the found values of $\sigma,$ both resonant and non-resonant, are typical of the scattering cross-sections of charge carriers on silicon-nitrogen impurities in graphene \cite{ervasti2015}. The present authors though are unaware of similar data for the trapping cross-sections on carbon impurities in hBN. It should be recalled also that the above consideration assumes independent traps, and that allowing for the interference effects may largely change the picture. It is known, for instance, that a sufficiently dense set of active centres is able to trap particles even if each centre separately is not. With these reservations, the following conclusion regarding the origin of $1/f$ noise in the considered heterostructure can be made. Taking into account that the noise level therein is near the quantum bound, the fact that the intricate behavior of the frequency exponent is so naturally explained as a result of the charge carrier trapping strongly suggests that what was observed in Ref.~\cite{das2021} is the {\it fundamental} $1/f$ noise.

\acknowledgments{The authors are grateful to S.~R.~Das and A.~K.~Behera for discussions of Ref.~\cite{das2021}, and to O.~V.~Pavlovsky (MSU) for useful comments. The study was conducted under the state assignment of Lomonosov Moscow State University.}


\begin{thebibliography}{}

\bibitem{pal2009}
A.~N.~Pal and A.~Ghosh, {\it Phys. Rev. Lett.} {\bf 102}, 126805 (2009).

\bibitem{liu2009}
G.~Liu, W.~Stillman, S.~Rumyantsev, Q.~Shao, M.~Shur, and A.~A.~Balandin, {\it Appl. Phys. Lett.} {\bf 95}, 033103 (2009).

\bibitem{lin2008}
Y.-M.~Lin and Ph.~Avouris, {\it Nano Lett.} {\bf 8}, 2119 (2008).

\bibitem{xu2010}
G.~Xu, C.~M.~Torres, Jr., Y.~Zhang, F.~Liu, E.~B.~Song,
M.~Wang, Y.~Zhou, C.~Zeng, and K.~L.~Wang,
{\it Nano Lett.} {\bf 10}, 3312 (2010).

\bibitem{pal2011}
A.~N.~Pal, S.~Ghatak, V.~Kochat, E.~Sneha, A.~Sampathkumar,
S.~Raghavan, and A.~Ghosh, {\it ACS Nano} {\bf 5}, 2075 (2011).

\bibitem{kumar2015}
M.~Kumar, A.~Laitinen, D.~Cox, and P.~J.~Hakonen,
{\it Appl. Phys. Lett.} {\bf 106}, 263505 (2015).

\bibitem{kumar2016}
Ch.~Kumar, M.~Kuiri, J.~Jung, T.~Das, and A.~Das,
{\it Nano Lett.} {\bf 16}, 1042 (2016).

\bibitem{cheng2010}
Z.~Cheng, Q.~Li, Z.~Li, Q.~Zhou, and Y.~Fang,
{\it Nano Lett.} {\bf 10}, 1864 (2010).

\bibitem{kakkar2020}
S.~Kakkar, P.~Karnatak, Md. Ali Aamir, K.~Watanabe, T.~Taniguchi, and A.~Ghosh, {\it Nanoscale} {\bf 12}, 17762 (2020).

\bibitem{kayyalha2015}
M.~Kayyalha and Y.~P.~Chen, {\it Appl. Phys. Lett.} {\bf 107}, 113101 (2015).

\bibitem{karnatak2016}
P.~Karnatak, T.~Phanindra Sai, S.~Goswami, S.~Ghatak, S.~Kaushal, and A.~Ghosh, {\it Nat. Commun.} {\bf 7}, 13703 (2016).

\bibitem{das2021}
A.~K.~Behera, Ch.~Th.~Harris, D.~V.~Pete, C.~J.~Delker, P.~E.~Vullum, M.~B.~Muniz, O.~Koybasi, T.~Taniguchi, K.~Watanabe, B.~D.~Belle, and S.~R.~Das, {\it ACS Applied Electronic Materials}  {\bf 3}(9), 4126 (2021).

\bibitem{rollin1953}
B.~V.~Rollin and I.~M.~Templeton, {\it Proc. Phys. Soc. B} {\bf 66}, 259 (1953).

\bibitem{caloyannides}
M.~F.~Caloyannides, {\it J. Appl. Phys.} {\bf 45}, 307 (1974).

\bibitem{kazakov2020}
K.~A.~Kazakov, {\it Phys. Lett. A} {\bf 384}, 126812 (2020).

\bibitem{kazakov2024}
K.~A.~Kazakov, {\it Int. J. Mod. Phys. B} {\bf 38}, 2450138 (2024).

\bibitem{mahan1972}
G.~Mahan, in {\it Polarons in Ionic Crystals and Polar Semiconductors}, ed. J.~T.~Devreese (North-Holland, Amsterdam, 1972), p. 553.

\bibitem{hooge1969}
F.~N.~Hooge, Phys.~Lett A {\bf 29}, 139 (1969).

\bibitem{stolyarov2015}
M.~A.~Stolyarov, G.~Liu, S.~L.~Rumyantsev, M.~Shur, A.~A.~Balandin,  {\it Appl. Phys. Lett.} {\bf 107}, 023106 (2015).

\bibitem{ervasti2015}
M.~M.~Ervasti, Zh.~Fan, A.~Uppstu, A.~V.~Krasheninnikov, and A.~Harju, {\it Phys. Rev.} B {\bf 92}, 235412 (2015).


\end{thebibliography}
\end{document}